\documentclass[journal,onecolumn,11 pt]{IEEEtran}

\usepackage{graphicx}
\usepackage{amsmath,epsfig,amssymb}
\usepackage{cite}

\begin{document}
\baselineskip=1cm
\title{Capacity Bounds of Finite Dimensional CDMA Systems with Fading/Near-Far Effects and Power Control}
\author{P. Kabir,~\IEEEmembership{Student Member,~IEEE}, M. H. Shafinia,~\IEEEmembership{Student Member,~IEEE},  F. Marvasti,~\IEEEmembership{Senior Member,~IEEE} , P. Pad,~\IEEEmembership{Student Member,~IEEE}
\thanks{Manuscript received}
\thanks{P. Kabir, M. H. Shafinia, F. Marvasti and P. Pad  are affiliated with ACRI and Electrical Engineering Department, Sharif University of Technology, Tehran, Iran. }}
\maketitle
\newtheorem{remark}{\textbf{Remark}}
\newtheorem{theorem}{\textbf{Theorem}}
\newtheorem{definition}{\textbf{Definition}}
\newtheorem{example}{\textbf{Example}}
\newtheorem{corollary}{\textbf{Corollary}}
\newtheorem{lemma}{\textbf{Lemma}}
\newtheorem{note}{\textbf{Note}}
\newtheorem{preposition}{\textbf{Preposition}}
\begin{abstract}
This paper deals with fading and/or near-far effects with or without power control on the evaluation of the sum capacity of finite dimensional Code Division Multiple Access (CDMA) systems for binary and finite non-binary inputs and signature matrices. Important results of this paper are that the knowledge of the received power variations due to input power differences, fading and/or near-far effects can significantly improve the sum capacity. Also traditional power controls can not improve the sum capacity; for the asymptotic case, any type of power control on the near-far effects is equivalent to the case without any power control. Moreover, for the asymptotic case, we have developed a method that determines bounds for the fading/near-far sum capacity with imperfect power estimation from the actual sum capacity of a CDMA system with perfect power estimation. To show the power and utility of the results, a number of sum capacity bounds for special cases are numerically evaluated.
\end{abstract}
\section{Introduction}\label{sec:intro}

In a CDMA system, each user is assigned a signature vector to transmit its data through a common channel. Different users have different distances from the receiver; thus, the received signals do not have the same power at the receiver end. Flat fading due to multi-path can also create power variations at the receiver end for different users.
In the present paper, we develop a common analytical tool for the evaluation of the sum capacity bounds for finite CDMA systems with fading/near-far effects with/without power control. In addition, we evaluate the asymptotic sum capacity bounds with imperfect power estimation from the sum capacity evaluations of Tanaka \cite{Tanaka} and Guo-Verdu \cite{Verdu4} with perfect power estimation for finite CDMA systems.


In the absence of near-far effects, the channel capacity has been evaluated for real and complex inputs \cite{Verdu1} and \cite{Verdu2}. However, for overloaded CDMA systems with finite input alphabets, only lower and upper bounds have been evaluated \cite{PadMarvastiConf,PadMarvasti,KasraVahid,Partthree}; a recent review of these papers is published in \cite{Javid}. Asymptotic results for finite input sum capacity have been derived by \cite{Tanaka} and \cite{Verdu4}. These asymptotic results are based on replica theory that has been proven rigorously for special cases \cite{TanakaProof}. The asymptotic results discussed in \cite{Verdu4} also cover the near-far effects with perfect power estimation.

For finite dimensional CDMA systems, we derive bounds for fading/near-far effects with or without power control. In this derivation, we will show that if we have perfect power estimation at the receiver end, the lower bound for sum capacity is actually increased. This implies that even in the absence of fading/near-far effects, if we allocate random powers to different users, we can actually improve the sum capacity. Also we will show that the traditional power control technique is surprisingly worse than no power control. Also, we shall see that water filling power control seems to be optimum for even binary input systems, although water filling power control is optimal for only Gaussian input signals\cite{macfad_1,macfad_2}.

For the asymptotic case, we derive bounds for the near-far CDMA systems in the absence of power estimation. Essentially, we derive a method that can estimate the sum capacity when perfect power estimation is not available. This method depends on the sum capacity evaluations of Tanaka\cite{Tanaka} and Guo-Verdu\cite{Verdu4} in the absence of near-far effects. 

At the transmitter side, there are several strategies to allocate powers to different users. In power control, the strategy is to compensate for the fading channel or the near-far effects. This type of power control tries to make the received powers from different users to be equal. This is not an optimum power policy. References \cite{macfad_1,macfad_2} have shown that for Gaussian real inputs, optimal power control is a generalization of water filling concept. Optimization of power control given a fading/near-far scenario for binary or finite size input alphabets are parts of future work. However, based on our numerical results, to be discussed later, water filling and even a random power allocation can increase the sum capacity.

The rest of the paper is organized as follows: In Section \ref{ChannelModel}, some mathematical preliminaries are given. In Section \ref{sec:MainIdea}, we will derive lower and upper bounds for the sum channel capacity for finite CDMA systems. In Section \ref{sec:Infinite}, we will develop lower and upper bounds for asymptotic CDMA systems with imperfect power estimation. Numerical computations of these bounds for special cases are in Section \ref{sec:simresult}. The conclusion and future works are covered in Section \ref{Sec:Conclusion}.\\

\section{Preliminaries}\label{ChannelModel}
In a DS-CDMA system, each user is assigned a signature vector. Each user multiplies its signature by its data, possibly with different powers,  and transmits it through a common channel. All vectors are added up together in the channel and the resultant vector embedded in noise is received. In such a system, without perfect power control, the assumption of receiving equal powers from all transmitters is no longer valid. Thus, in a synchronous CDMA system with $n$ users and $m$ chips in the presence of noise and near-far effects, the channel model is
\begin{align} \label{Eq:NearFarModel} 
&Y=\sum_{i=1}^{n}\frac{1}{\sqrt{m}}A_i C_i M_i X_i +N=\nonumber\\&\frac{1}{\sqrt{m}}\mathbf{A} \mathbf{C}\mathbf{M}X+N=\frac{1}{\sqrt{m}}\mathbf{A} \mathbf{C}(\mathbf{G}+\mathbf{E})X+N,
\end{align} 
where $\mathbf{A}=[A_1|\cdots|A_n]\in {\mathcal S}^{m\times n}$ is the signature matrix and $\mathcal S$ is the set of signature alphabets. Also, $\mathbf{M}$ and $\mathbf{C}$, are diagonal matrices corresponding to channel gain/loss and power allocation, respectively. $X=[X_1,\cdots,X_n]^T\in{\mathcal I}^n$ is the user data vector, where $\mathcal{I}$ is the set of input alphabets. Here, $\mathcal I$ and $\mathcal S$ can be arbitrary binary, non-binary, real or complex numbers. In addition, $N=[N_1,\cdots,N_m]^T$ is i.i.d. Gaussian noise vector in which each entry has a variance of $\sigma^2$.  Also $M_i = G_i + E_i$, where $G_i$ is the estimation of the near-far or fading at the receiver and $E_i$ is the estimation error. We assume that $E_i$ , $G_i$ and $C_{i}$ are i.i.d. random variables with distribution $\mathcal N (0, \rho^{2})$, $g$, and $c$, respectively.


We define the Channel Estimation Ratio (CER) of a CDMA system as
\begin{eqnarray}\label{Eq:PCF}
\text{CER}_{\text{dB}}=10\log_{10}{\frac{\mathbb{ E}{\left[{\text{Re}(G_i)^2}\right]}}{\text{Var}{\left[{\text{Re}(E_i)}\right]}}},
\end{eqnarray}
where $\text{Re}(\cdot)$ is the real part of a complex function. CER is the ratio of the estimated channel power divided by the power estimation error.\\

Define $\mathcal{C}\left(m,n,\mathcal{I},\mathcal{S}_\pi,\eta,g,c,\rho\right)$ to be the sum capacity over all random matrices $A$ with independent components of distribution $\pi(\cdot)$, where $\pi(\cdot)$ is the distribution on $\mathcal S$. The normalized Signal to Noise Ratio (SNR), $\eta$, is defined by $\eta =\frac{{\sigma_p^2}{\left(\sigma_\pi^2+\mu_\pi^2\right)}{\left(\sigma_c^2+\mu_c^2\right)}}{2\sigma^2} $, where $\mu$ and $\sigma$ are the mean and the variance of the random variables. Also, we assume that $\mu_p=0$, where $p(\cdot)$ is the probability distribution function on $\mathcal I$.

\section{Bounds for the Finite Dimensional CDMA System with Perfect Power Estimation}\label{sec:MainIdea}
In this section, we will derive lower and upper bounds for capacity of finite dimensional CDMA systems with fading or near-far effects. Throughout this section we assume that perfect power estimation is available, i.e., $\rho=0$.
First, we propose a general lower bound for $\mathcal{C}\left(m,n,\mathcal{I},\mathcal{S}_\pi,\eta,g,c,\rho=0\right)$.
~\\
\begin{theorem}\label{Th:MainTheorem}
\textbf{Lower Bound for Finite Dimensional CDMA Systems with Perfect Power Estimation}\\
For the above channel model, when power estimation is perfect, we have
\begin{align}\label{Eq:global}
&\mathcal{C}\left(m,n,\mathcal{I},\mathcal{S}_\pi,\eta,g,c,\rho=0\right)\geq\mathbb E_{\mathbf G,\mathbf C}\Bigg\{\sup_{p}\sup_{\gamma}\Big\{-m(\gamma\log{e}\nonumber\\&-\log{\left(1+\gamma\right)})
-\log{\mathbb E_{\tilde{X}}\left(\left(\mathbb E_{b}\left(e^{\frac{-\gamma r^2}{2\left(1+\gamma\right)m}|b^T\mathbf {CG}\tilde X|^2}\right)\right)^m\right)}\Big\}\Bigg\},
\end{align}
\end{theorem}
where $r = \sqrt{\frac{2\eta}{{\sigma_p^2}{\left(\sigma_\pi^2+\mu_\pi^2\right)}{\left(\sigma_c^2+\mu_c^2\right)}}}$; $b$ and $\tilde X$ are, respectively, vectors of length $n$ with i.i.d. entries of distribution $\pi(\cdot)$ and $\tilde p(\cdot)$, in which $\tilde p(\cdot)$ is defined to be the probability law on $\tilde{ \mathcal I} = \mathcal I - \mathcal I$, which is the difference of two independent random variables of pdf $p(\cdot)$.

For the proof, refer to Appendix \ref{ProofMainTheorem}.
Now, we propose a general upper bound for the finite case.
~\\
\begin{theorem}\label{Th:FiniteUpper}
\textbf{Upper Bound for Finite Dimensional CDMA Systems with Perfect Power Estimation}\\
For the above channel model, when power estimation is perfect, we have
\begin{align}
&\mathcal{C}\left(m,n,\mathcal{I},\mathcal{S}_\pi,\eta,g,c,\rho=0\right)\leq\nonumber\\ &min \Bigg(m \log |\mathcal I|,\mathbb E_{\mathbf {C,G},A^1} \Big\{ m(\mathbb H(Y_1)-\mathbb H(N_1))\Big\}\Bigg),
\end{align}
\end{theorem}
where $Y_1 = \frac{1}{\sqrt{m}}\sum_{i=1}^na_iX_iC_iG_i+N_1$ and $A^1=[a_1,\dots,a_n]$ is the first row of matrix A.  

The proof is presented in Appendix \ref{prooffiniteupper}.
In the next section we will use these bounds to find lower and upper bounds for asymptotic CDMA systems.

\section{Asymptotic Bounds for the Infinite Dimensional CDMA Systems with Imperfect Power Estimation}\label{sec:Infinite}
For the asymptotic case, we need to define the sum capacity as given by 
\begin{equation}
\zeta\left(\beta, \mathcal I, \mathcal S_\pi, \eta, g,c,\rho\right)=\lim_{\begin{smallmatrix}m,n\rightarrow\infty\\ n/m\rightarrow\beta\end{smallmatrix}}{\mathcal{C}\left(m,n,\mathcal{I},\mathcal{S}_\pi,\eta,g,c,\rho\right),}
\end{equation}
where $\beta = \frac{n}{m}$ is the loading factor.

Asymptotic results for binary CDMA is developed by Tanaka\cite{Tanaka}. Guo-Verdu's paper \cite{Verdu4} has extended Tanaka's results to non-binary finite size alphabet with near-far effects with perfect power estimation, including various optimal and suboptimal detectors. In this section, we plan to obtain bounds for sum capacity when perfect power estimation is not available ($\rho\neq 0$). Below, we will give examples where we can find bounds for Tanaka and Guo-Verdu results when perfect power estimation is not available. Throughout this section we assume that $\mathbf C = \mathbf I$, which means that there is no power allocation mechanism.
\begin{theorem}\label{Th:ImperfectReal}
\textbf{Lower and Upper Bounds for the Sum Capacity of CDMA Systems with Imperfect Power Estimation}\\
Suppose that $\rho$ is not zero, which implies that we have an imperfect estimation of user powers. We then have
\begin{align}
\zeta(\beta,\{\pm 1\} ,\mathcal S_\pi,\eta ,g,\delta\left(\cdot-1\right),\rho)\geq \zeta(\beta,\{\pm 1\},\mathcal S_\pi,\eta_l ,g,\delta\left(\cdot-1\right),\rho=0),
\end{align}
\begin{align}
\zeta(\beta,\{\pm 1\},\mathcal S_\pi,\eta ,g,\delta\left(\cdot-1\right),\rho)\leq \zeta(\beta,\{\pm 1\},\mathcal S_\pi,\eta_u ,g,\delta\left(\cdot-1\right),\rho=0),
\end{align}
where $c(\cdot)=\delta\left(\cdot-1\right)$ is the Dirac delta function representing the pdf of a point process at point $1$ and
\begin{align}
\eta_l = \frac{\eta}{1+{\xi\rho^2}\eta\left(1+\sqrt \beta\right)^2},\nonumber\\\eta_u = \frac{\eta}{1+{\xi\rho^2}\eta\left(1-\sqrt \beta\right)^2}
\end{align}
\end{theorem}
~\\
in which, $\xi=1$ for real CDMA systems and $\xi=2$ for complex CDMA systems.
For the proof, please refer to Appendix \ref{ProofLowerUpperImperfect}.
~\\
\begin{example}\label{example:RealChannel}
\textbf{Tanaka's Capacity}\\
For the binary input vectors and signature matrices, from Theorem \ref{Th:ImperfectReal} and \cite{Tanaka}, the sum capacity is bounded by 
\begin{eqnarray}
\nonumber
&\frac{1}{2\beta}\log{\left(1+{2\eta_u\beta\left(1-\theta\right)}\right)}+q\left(\lambda,\theta\right)\log\left(e\right)
\end{eqnarray}
\begin{eqnarray}\label{Eq:Tanaka}\nonumber
&\geq \zeta(\beta,\left\{\pm1\right\},\left\{\pm1\right\},\eta,\delta\left(\cdot-1\right),\rho)\geq
\end{eqnarray}
\begin{eqnarray}\label{fourteen}
&\frac{1}{2\beta}\log{\left(1+{2\eta_l\beta\left(1-\theta\right)}\right)}+q\left(\lambda,\theta\right)\log\left(e\right),
\end{eqnarray}
where $q\left(\lambda,\theta\right)$ is defined by
\begin{eqnarray}
q\left(\lambda,\theta\right)=\frac{\lambda}{2}\left(1+\theta\right)-\int\ln{\left(\cosh\left(\sqrt{\lambda}Z+\lambda\right)\right)}D_Z,
\end{eqnarray}
in which $D_Z$ is the standard normal measure,
\begin{align}
\lambda=\frac{1}{\sigma_l^2+\beta\left(1-\theta\right)},~~~~
\theta=\int\tanh\left(\sqrt{\lambda}Z+\lambda\right)D_Z.
\end{align}
This result is based on replica theory, a rigorous proof of \cite{Tanaka} is given in \cite{TanakaProof} for $\beta\leq\alpha_s\approx 1.49$.\\
\end{example}
\begin{example}\label{example:GuoVerdu}
\textbf{Guo-Verdu's Capacity}\\
For the binary input vectors and signature matrices, Guo and Verdu\cite{Verdu4} derive the channel capacity in the asymptotic case when the power estimation is perfect,
\begin{align}\nonumber
\zeta\left(\beta,\mathcal{I},\mathcal{S}_\pi,\eta,g,c,\rho=0\right)=\\\nonumber \beta \text{E}\{-\int \frac{e^{\frac{z^2}{2}}}{\sqrt{2\pi}}\log \cosh\left(\psi snr-z\sqrt{\psi snr}\right)\text{d}z \}\\+\beta\psi\text{ E}(snr) \log e+\frac{1}{2}\left[(\psi-1)\log e - \log \psi\right],
\end{align}

where $\psi$ is the solution of this fixed point equation:
\begin{align}
\frac{1}{\psi}=1+\beta\text{E}\{snr\left[1-\int\frac{e^{-\frac{z^2}{2}}}{\sqrt{2\pi}}\tanh\left(\psi snr-z\sqrt(\psi snr)\right)\text{ dz}\right]\},
\end{align}
The expectations is over $snr=\frac{n}{m}\times\frac{\eta}{\mu_c^2+\sigma_c^2}({G_{1}}^2+{C_{1}}^2)$, where $G_1$ and $C_1$ are the random variables with probability distribution functions $g$ and $c$, respectively. 
\end{example}


\section{Numerical Results}\label{sec:simresult}
In this section we try to numerically evaluate the bounds derived in the previous sections. For finite CDMA systems because of computational complexity, the simulation results are for small values of $m$ and $n$, namely, ($m=5$, $n=10$) and ($m=3$, $n=9$). For asymptotic results, we have used $\beta=2,3$. 

\begin{figure}[t]
\centering
\includegraphics[width=19cm]{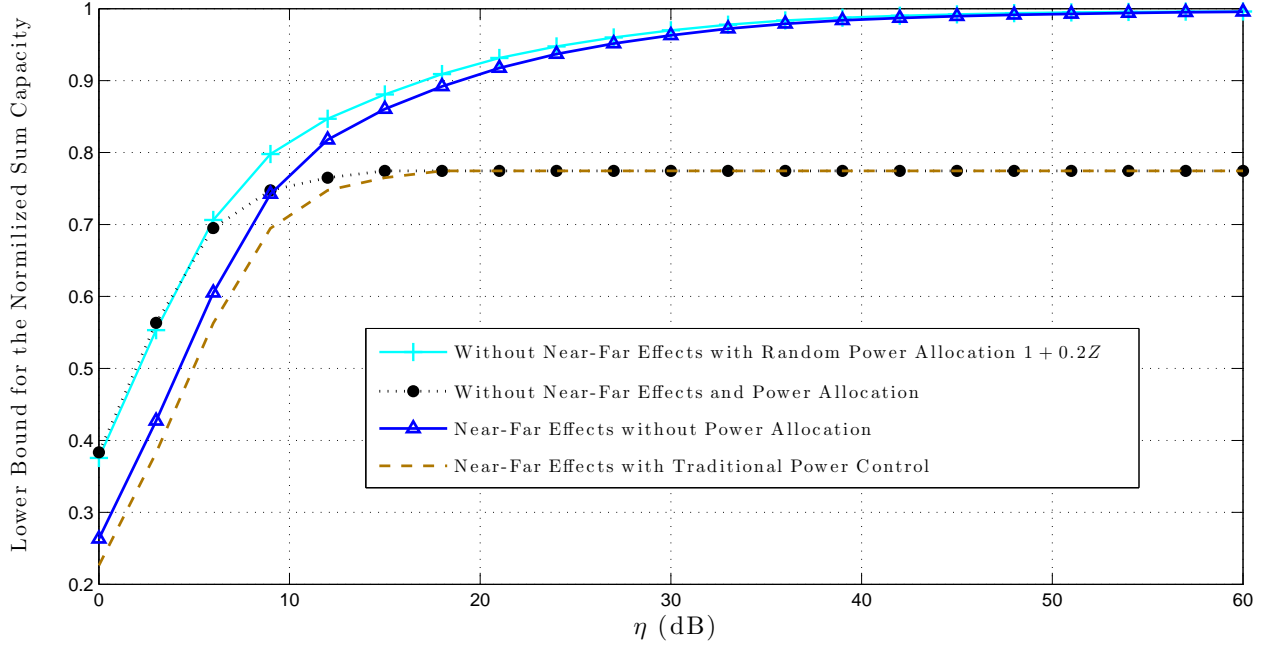}
\caption{Sum capacity bounds for binary CDMA systems with m=5 and n=10, where $G_{i}$  the uniform distribution between 0.5 and 1, also $Z\sim \mathcal{N}(0,1)$..}
\label{Fig: FigFinite510}
\end{figure}

Figure \ref{Fig: FigFinite510} shows a comparison of the sum capacity lower bound versus the normalized SNR ($\eta$) for four different scenarios. The dotted line in the figure in the lower bound for the sum capacity for a binary CDMA system without any near-far effects. The solid curve is the lower bound for the sum capacity with near-far effects when we have perfect knowledge about the power at the receiver. The dashed curve is for the case when we have the traditional power compensation for near-far effects with perfect power estimation. Pale solid line is the sum capacity lower bound with random power allocation. These numerical results are for a CDMA system with dimensions of (m=5, n=10) and hence an overloading factor $\beta=2$. The results in this figure are quiet surprising for the normalized SNR values of $\eta$ greater than 9 dB; a CDMA system with near-far effects has a higher lower bound for the sum capacity than that of a case without any near-far effects when we have perfect knowledge about the variation of power at the receiver end\footnote{In power allocation schemes, assigning the same power to all users might not be the best strategy. Thus, we expect the existence of power allocation schemes that improve the sum capacity. Our lower bounds confirm this proposition. Although the improvement of lower bounds do not mean that the actual capacity has increased, since the sum capacity is defined as the supremum over all input powers and distributions and signature matrices, the sum capacity can only improve.}. The second surprise is that the traditional power compensation for the near-far effects is worse than the near-far effects without any compensation when we have perfect knowledge about the received power. This implies that, equivalently, if we have random power allocation at the transmitter side, we can improve the lower bound for the sum capacity. Essentially, the solid curve in Fig. \ref{Fig: FigFinite510} is shifted to the left by a fixed amount as explained in footnote\footnote{$\eta$ is normalized by the transmitter power, this implies that in power control and power allocation strategies we have to consider the average power variations in the normalized $\eta$. However, in fading or near-far effects, the parameter $\eta$ is not changed; this creates a shift in the curve.}.

\begin{figure}[t]
\centering
\includegraphics[width=19cm]{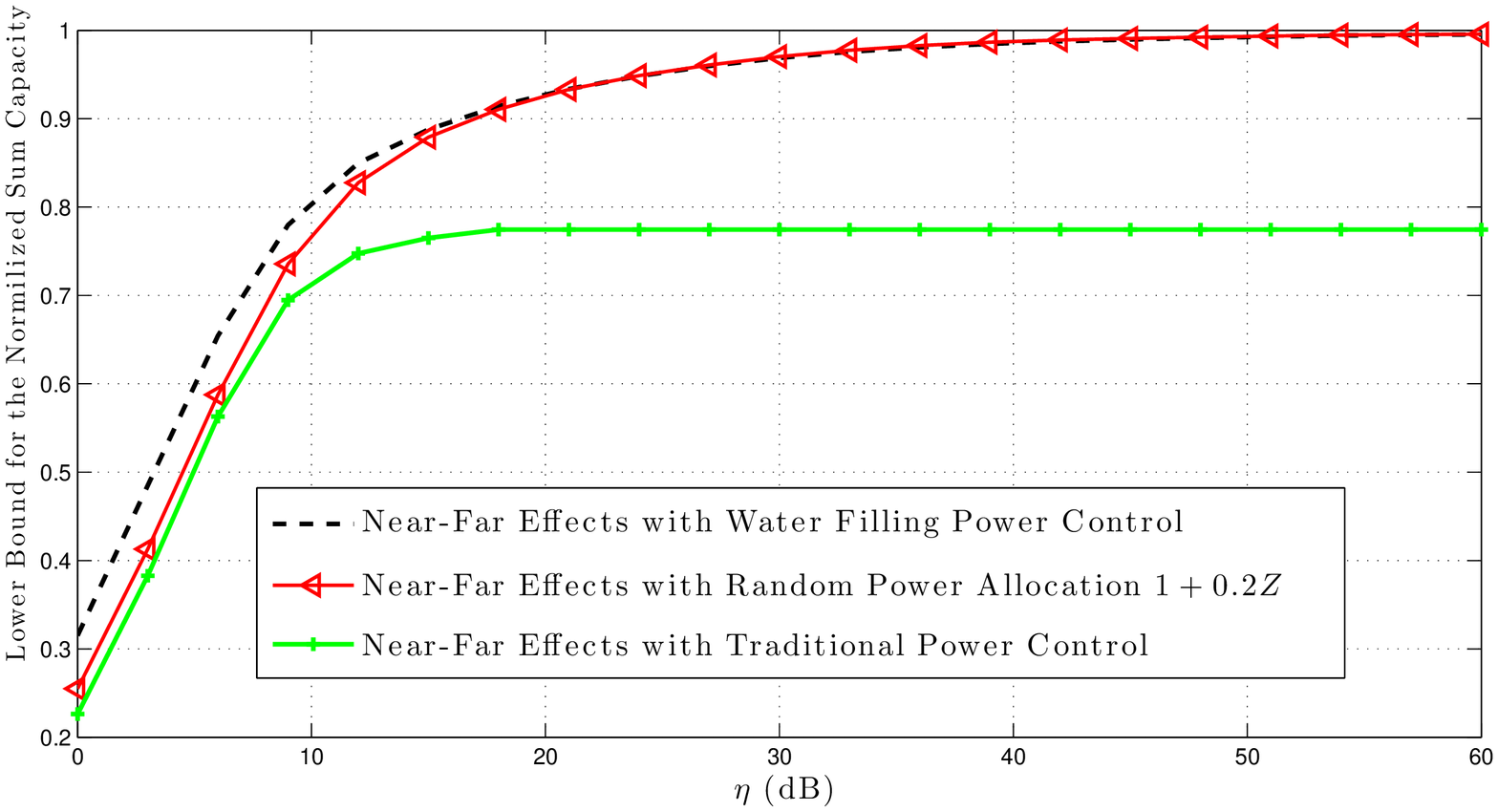}
\caption{Sum capacity bounds for binary CDMA systems with m=5 and n=10, where $G_{i}$  the uniform distribution between 0.5 and 1, also $Z\sim \mathcal{N}(0,1)$.}
\label{Fig: Finite510}
\end{figure}

Figure \ref{Fig: Finite510} shows a comparison of three strategies for power control. The lower solid line in this figure is the same as the lower curve in Fig. \ref{Fig: FigFinite510} where the near-far effects are compensated by traditional power control. The middle solid curve shows the lower bound for the sum capacity when a Gaussian random power control is used. The dotted curve in this figure shows power control using water filling. Clearly, the water filling power control has the best performance. However, an appropriate random strategy for power control can approach the water filling strategy. The lower bounds for the sum capacity for the water filling and the random power control are approaching 1 for high values of $\eta$; this implies that the lower bounds become very tight for the noiseless case.

\begin{figure}[t]
\centering
\includegraphics[width=19cm]{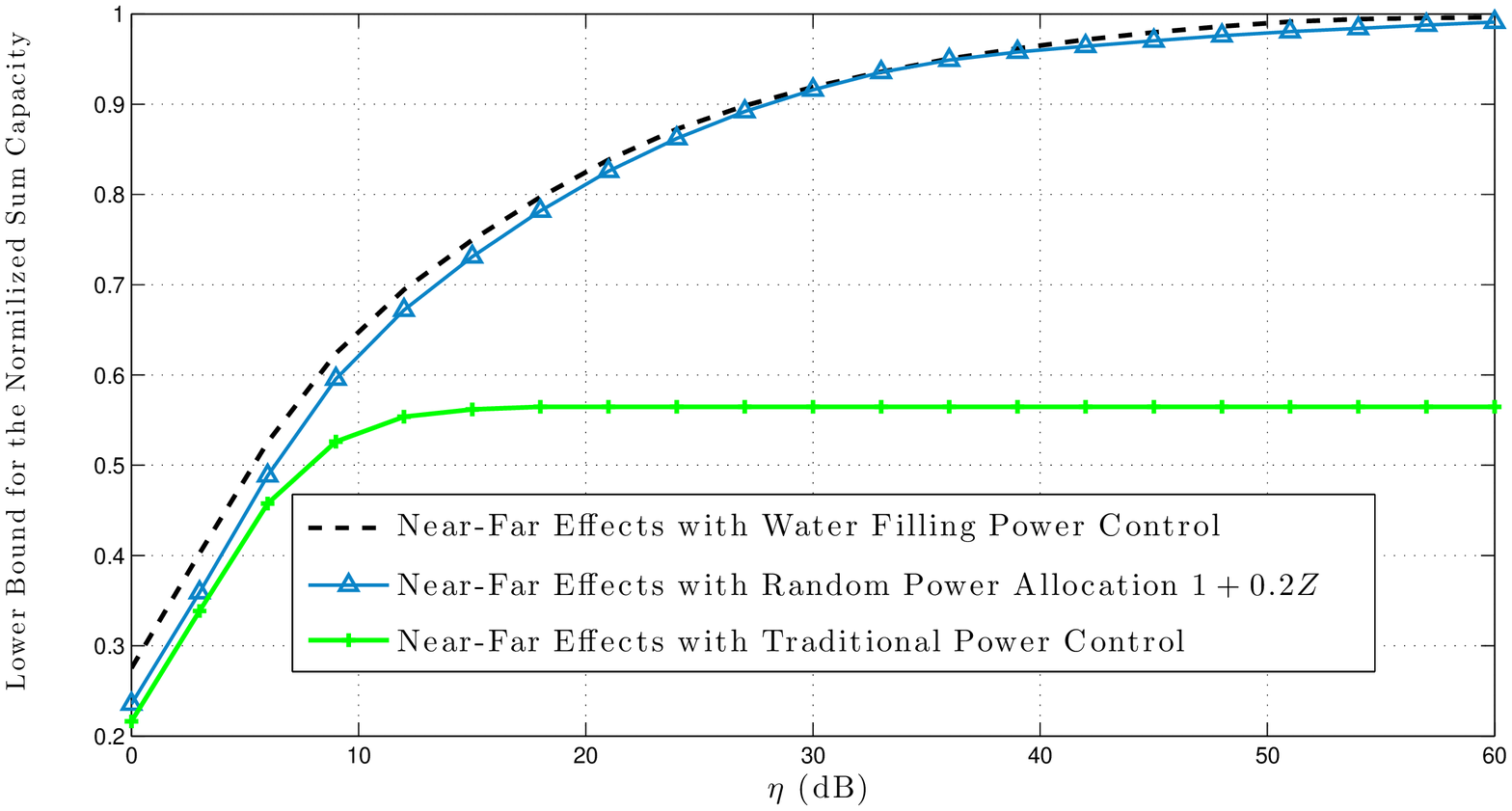}
\caption{Sum capacity bounds for binary CDMA systems with m=3 and n=9, where $G_{i}$ has uniform distribution between 0.5 and 1, also $Z\sim \mathcal{N}(0,1)$.}
\label{Fig: Finite39}
\end{figure}

Figure \ref{Fig: Finite39} is similar to the previous figure except that the dimension of the CDMA system is (m=3, n=9) and hence $\beta=3$. The same observations as Fig. \ref{Fig: Finite510} also hold.

\begin{figure}[t]
\centering
\includegraphics[width=19cm]{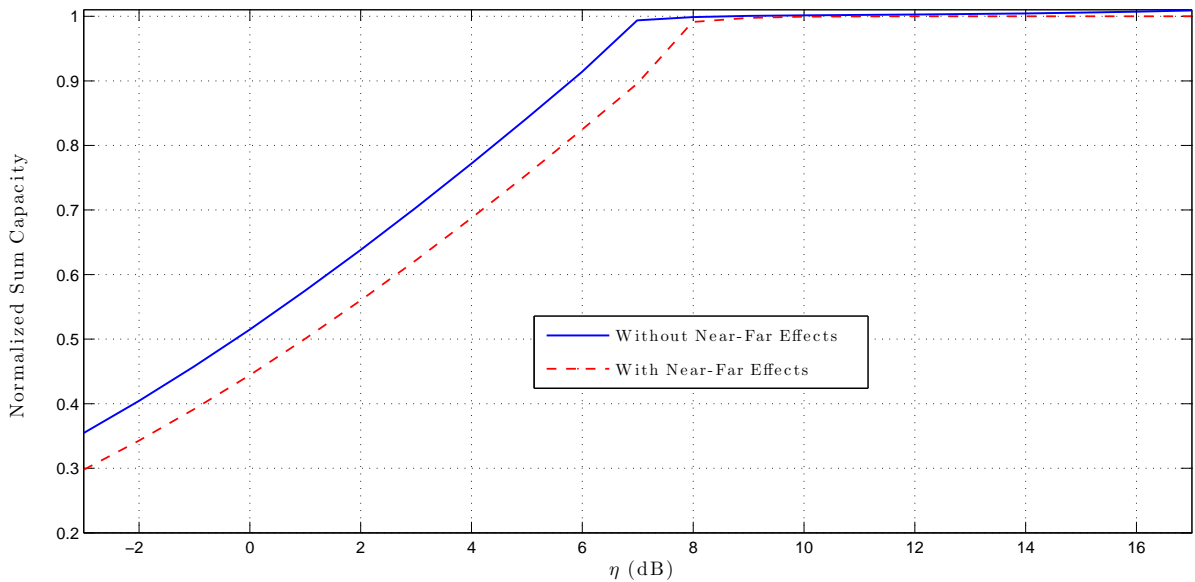}
\caption{Sum capacity bounds for asymptotic binary CDMA systems with $\beta=2$, where $G_{i}$ has uniform distribution between 0.5 and 1.}
\label{Fig:HosseinAssymp}
\end{figure}

On the other hand, the asymptotic behaviors are drastically different. In the absence of near-far effects, any power allocation does not change the lower bound. Also, in the presence of new-far/fading channels, no types of power compensation can improve the results. Essentially, we have two lower bounds as shown in Fig. \ref{Fig:HosseinAssymp}. This figure is the numerical results of Tanaka \cite{Tanaka} and Guo-Verdu \cite{Verdu4} sum capacities with or without near-far effects.

\begin{figure}[t]
\centering
\includegraphics[width=19cm]{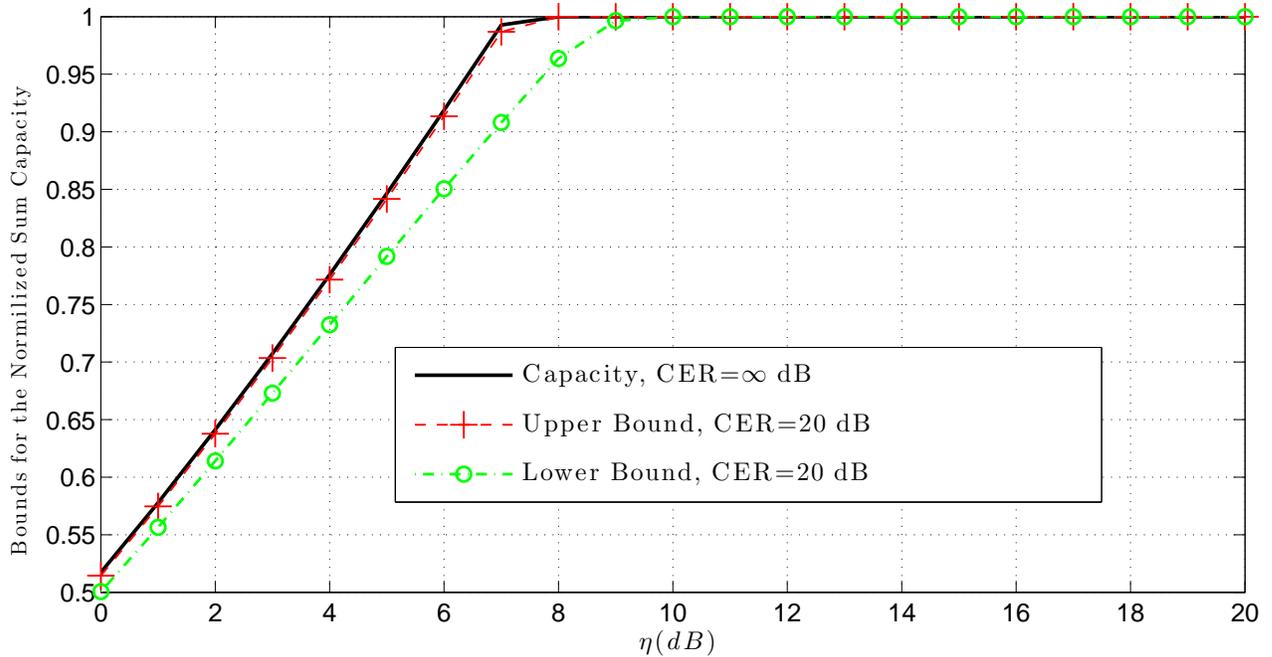}
\caption{Sum capacity bounds for binary CDMA systems with $\beta=2$.}
\label{Fig:NF2_ICC}
\end{figure}

\begin{figure}[t]
\centering
\includegraphics[width=19cm]{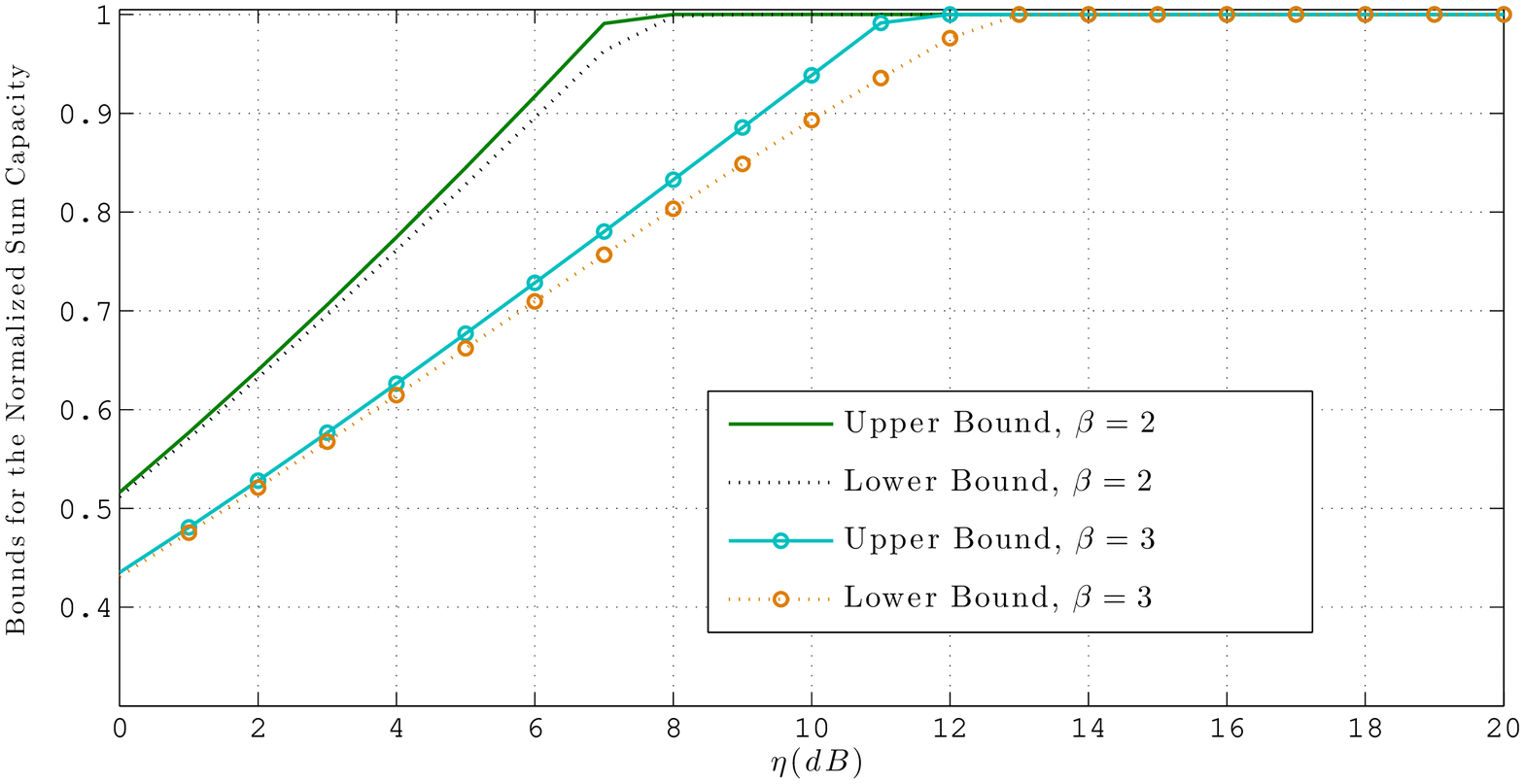}
\caption{A comparison between sum capacity bounds for binary CDMA systems with CER=$20dB$ for two values of $\beta$.}
\label{Fig:NF3_ICC}
\end{figure}

We have also evaluated the asymptotic sum capacity with imperfect power estimation. Figure \ref{Fig:NF2_ICC} shows a comparison among the actual normalized asymptotic sum capacity derived by Tanaka when there is no near-far effects with perfect power estimation and bounds for the equivalent CDMA system with near-far effects and imperfect power estimation (CER=$20dB$). This figure clearly shows the degradation of the sum channel capacity in near-far/fading environments. Figure \ref{Fig:NF3_ICC} gives a comparison of the asymptotic bounds for the sum capacity for two values of $\beta$ in the presence of near-far effects with a CER of $25dB$. This figure also shows the bounds are tight for high values of CER.

%
%
%

\section{Conclusion and Future Works}\label{Sec:Conclusion}

The first contribution of this paper is the development of lower and upper bounds for the sum channel capacity of finite dimensional CDMA systems with fading and/or near-far effects for binary and finite non-binary inputs. An interesting observation in the simulations of such finite dimensional systems is that when we have fading or near-far effects with perfect power estimation, the sum capacity of the "degraded" CDMA system is surprisingly greater than the ideal case for SNR values greater than a certain threshold. For the asymptotic case, the inequality in the previous system becomes equality.

Also, the effects of power allocation and power control on the sum capacity with near-far effects are studied. We have concluded that traditional power control can not improve the sum capacity unlike water filling and appropriate random power allocation methods. Again power control strategies can not improve the sum capacity for infinite dimensional systems.

Equivalently, we could consider random power allocation for the users at the transmitters without fading/near-far effects. In this case, the sum capacity is always greater than or equal to the case of equal power allocation for the finite dimensional CDMA. In the asymptotic case, the sum capacities coincide, which is also a surprise. 

The other contribution is the development of a method that translates a near-far sum capacity problem with imperfect power estimation to the evaluation of the capacity for a CDMA system with perfect power estimation.

For future work, we suggest to use a Markov chain for the power model which can improve the bounds. Also, optimal power allocation for binary and non-binary CDMA systems is an interesting subject.

\appendix

\subsection{Proof of Theorem \ref{Th:MainTheorem}, Lower Bound for Finite Dimensional CDMA Systems with Perfect Power Estimation}\label{ProofMainTheorem}

For a fixed $\mathbf{A}$, $\mathbf {G}$ and $\mathbf {C}$, we have
\begin{align}\nonumber
&\mathbb{I}(X;Y)=h(Y)-h(N)=\\\nonumber
&\mathbb{E}_{Y}{\left(-\log{f_{Y}}(Y)\right)}-\mathbb{E}_{N}{\left(-\log{f_N{(N)}}\right)}\\\nonumber
&=\mathbb{E}_{X,N}{\left(-\log\frac{f_{Y}{\left({\frac{1}{\sqrt{m}}\mathbf{A}\mathbf C\mathbf GX+N}\right)}}{f_N{\left(N\right)}}\right)}.
\end{align}
Using Jensen's inequality, we get the following :
\begin{align}\nonumber
&\mathbb{I}(X;Y)=\\\nonumber
&\mathbb{E}_N{\left(-\log{\left(\sum_{x,x'}{\mathbb{P}_X(x)\mathbb{P}_X(x')\frac{f_N{\left(\frac{1}{\sqrt{m}}\mathbf{ACG}(x'-x)+N\right)}}{f_N(N)}}\right)}\right)}.
\end{align}
Using Jensen's inequality, we get 

\begin{align}
&\mathbb{E}_{A}{\left(\mathbb{I}(X;Y)\right)}\geq -m\mathbb{E}{\left(q(N_1)\right)}-\\\nonumber&\log{\left(\sum_{x,x'}{\mathbb{P}_X(x)\mathbb{P}_X(x')\mathbb{E}_{N,A}{\left(\frac{f_N{\left(\frac{1}{\sqrt{m}}\mathbf A\mathbf{C G}(x'-x)+N\right)}}{2^{\sum{q(N_i)}}f_N(N)}\right)}}\right)}.
\end{align}
Since $f_N\left(x\right)=\prod_{j=1}^m{f_{N_j}\left(x_j\right)}=\prod_{j=1}^m{f\left(x_j\right)}$, we get
\begin{eqnarray}\nonumber
\frac{f_N\left(\frac{1}{\sqrt{m}}ACG\left( x'- x\right)+N\right)}{2^{\sum{q\left(N_i\right)}}f_N\left(N\right)}=\prod_{j=1}^m{\frac{f\left(\frac{1}{\sqrt{m}}ACG\left( x'- x\right)_j+N_j\right)}{2^{q\left(N_j\right)}f\left(N_j\right)}}.
\end{eqnarray}
For fixed $x$ and $x'$, $\mathbf A\mathbf C\mathbf G{\left(x'-x\right)}_j$ and $N_j$ are independent for $j=1,\dots,m$.\\
Thus, the expectation and the product operators  commute; we then get
\begin{align}
&\mathbb E_{N,A}\left(\frac{f_N\left(\frac{1}{\sqrt{m}}ACG\left( x'-x\right)+N\right)}{2^{\sum{q\left(N_i\right)}}f_N\left(N\right)}\right)=\nonumber\\&={\left(\mathbb E_{N_1,A}\left(\frac{f\left(\frac{1}{\sqrt{m}}ACG\left(x'-x\right)_1+N_1\right)}{2^{q\left(N_1\right)}f\left(N_1\right)}\right)\right)}^m.
\end{align}
After computation of $\mathbb E_{N_1}$ and making simplifications we get

\begin{align}\nonumber
&\mathbb{E}_{A}{\left(\mathbb{I}(X;Y)\right)}\geq\\
&=-m\mathbb{E}{\left(q(N_1)\right)}-\log{\left({\mathbb{E}_{\tilde{X}}{\left(\mathbb{E}_{N_1,a}{\left(2^{-q{\left(N_1-\frac{a^TCG.\tilde{X}}{\sqrt{m}}\right)}}\right)}\right)}^m}\right)}.
\end{align}
~\\
Now averaging over all $\mathbf G$ and $\mathbf C$, we arrive at the following lower bound
\begin{align}
&\mathcal{C}\left(m,n,\mathcal{I},\mathcal{S}_\pi,\eta,g,c,\rho=0\right)\geq\mathbb E_\mathbf {G,C}\Bigg\{\sup_{p}\sup_{\gamma}\Big\{-m(\gamma\log{e}\nonumber\\&-\log{\left(1+\gamma\right)})
-\log{\mathbb E_{\tilde{X}}\left(\left(\mathbb E_{b}\left(e^{\frac{-\gamma r^2}{2\left(1+\gamma\right)m}|b^T\mathbf C\mathbf G\tilde X|^2}\right)\right)^m\right)}\Big\}\Bigg\},
\end{align}

$\Box$.

\subsection{Proof of Theorem \ref{Th:FiniteUpper}, Upper Bound for Finite Dimensional CDMA Systems with Perfect Power Estimation}\label{prooffiniteupper}
For the proof, it can be seen that the first term is trivial.
The second part operates as follows: 
\begin{eqnarray}\nonumber
&\mathbb I (X;Y)= \mathbb H(Y)-\mathbb H(Y|X) = \mathbb H(Y)-\mathbb H(N)\leq\\&\Sigma_{i=1}^m\mathbb H(Y_i)-m\mathbb H(N_1).
\end{eqnarray}
Due to symmetry, it is easy to show that $\mathbb E_A\left\{\mathbb H(Y_i)\right\}$ is the same for all $1\leq i\leq m$ and is equal to $\mathbb E_{A^1}\left\{\mathbb H(Y_1)\right\}$, where $A^1$ denotes the first row of $A$; thus,
\begin{eqnarray}\nonumber
&  \mathbb E_ A\Big\{ \frac{1}{n}\{\mathbf \max_{p(x)}\mathbb I(X;Y)\} \Big\} \leq\\& \mathbb E_{A^1}\Big\{ \frac{1}{\beta}(\mathbb H(Y_1)-\mathbb H(N_1))\Big\}.
\end{eqnarray}

\subsection{Proof of Theorem \ref{Th:ImperfectReal}, Lower and Upper Bounds for the Sum Capacity of CDMA Systems with Imperfect Power Estimation}\label{ProofLowerUpperImperfect}
For a real CDMA system, the system model in (\ref{Eq:NearFarModel}) can be written as
\begin{eqnarray}\label{Eq:NewNoi}
Y=\frac{1}{\sqrt{m}}\mathbf{AG}X+\Big(\frac{1}{\sqrt{m}}\mathbf{AE}X+N\Big).
\end{eqnarray}
Assume that $\mathcal{I}= \left\{\pm1\right\}$, then the entries of $E X$ are i.i.d Gaussian random variables of variance $\rho^2$ independent of entries of $G X$. Suppose that the minimum and maximum eigenvalues of $\frac{1}{m}\mathbf{A}\mathbf{A}^*$ are $\lambda_{min}$ and $\lambda_{max}$, respectively. If $\frac{1}{\sqrt{m}}\mathbf{A}EX+N$ are substituted with $\left(\sqrt{\lambda_{min}\rho^2+\sigma^2}\right)W$ and $\left(\sqrt{\lambda_{max}\rho^2+\sigma^2}\right)W$ where $W$ is a standard Gaussian vector, two systems, with a capacity greater and less than the system represented in (\ref{Eq:NewNoi}) are obtained.\\
Since entries of matrix $\mathbf{A}$ are chosen independently at random from a set $\mathcal{S}$ from a distribution $\pi\left(\cdot\right)$, with $\mu_{\pi}=0$ and  $m,n\rightarrow\infty$ such that $n/m\rightarrow\beta$, Then by using the Marcenko-Pastur theorem \cite{MarcenkoPastur}, the following equations are obtained:
\begin{eqnarray}
\mathbb{P}\left(\lambda_{min}\geq\sigma_{\pi}^2\left(\sqrt{\beta}-1\right)^2\right)\rightarrow 1,
\end{eqnarray}
\begin{eqnarray}
\mathbb{P}\left(\lambda_{max}\leq\sigma_{\pi}^2\left(\sqrt{\beta}+1\right)^2\right)\rightarrow 1.
\end{eqnarray}
Therefore, by utilizing the proposed lower and upper bounds for CDMA systems with perfect power control, it is possible to achieve lower and upper bounds for CDMA systems with near-far effects.
Note that when $\rho=0$, these formulas yield perfect power estimation formulas. Also, the complex bound is derived in the same manner.\hfill$\Box$

\subsection{Lower Bound for the Sum Capacity of CDMA Systems with Perfect Power Estimation for Real and Complex CDMA Systems}\label{ProofAsyLowNearFarCSIReal}
For perfect power estimation, $\rho=0$ and hence the user amplitudes are known without any ambiguity at the receiver. The following two bounds are related to lower and upper bounds for the sum capacity of the CDMA systems with near-far effects.
In a CDMA system with perfect power estimation, we have the following lower bound for the average sum capacity
\begin{align}
&\zeta(\beta,\mathcal I,\mathcal S_\pi,\eta ,g,c,\rho=0)\geq \sup_{p}\sup_{\gamma}
\Big\{-\frac{1}{2\beta}(\gamma\log{e}-\log{\left(1+\gamma\right)})\nonumber\\&-\log{e}\times\nonumber\\&\sup_{\theta\in \mathbb R}\{-\frac{1}{2\beta}\ln\Big(1+\frac{2\eta\gamma\beta\theta}{\sigma_p^2(1+\gamma)(\sigma_c^2+\mu_c^2)}\Big)-I(\theta)\}\Big\},
\end{align}
where $I(\theta)=\sup_{x\in \mathbb R}\{\theta x-\ln\mathbb E(e^{x(\tilde X_1C_1G_1)^2})\}$ is the Legendre transform of $(\tilde X_1C_1G_1)^2$, in which $C_1$ and $G_{1}$ are as defined in Section \ref{ChannelModel} and $\tilde X$ is the difference random variable as defined previously.

From Theorem \ref{Th:MainTheorem}, by changing the order of $\mathbb E_\mathbf{C,G}$ and $sup$ and applying Jensen's inequality for $\log$ function, we get\\
\begin{align}\label{Eq:global1}
&\mathcal{C}\left(m,n,\mathcal{I},\mathcal{S}_\pi,\eta,g,c,\rho=0\right)\geq\sup_{p}\sup_{\gamma}\Big\{-m(\gamma\log{e}\nonumber\\&-\log{\left(1+\gamma\right)})
-\log{\mathbb E_{\tilde{X},\mathbf {C,G}}\left(\left(\mathbb E_{b}\left(e^{\frac{-\gamma r^2}{2\left(1+\gamma\right)m}|b^T\mathbf {CG}\tilde X|^2}\right)\right)^m\right)}\Big\},
\end{align}
Thus 
\begin{align}
&\zeta(\beta,\mathcal I,\mathcal S_{\pi},\eta,g,c,\rho=0) = \lim_{n\rightarrow\infty}\frac{1}{n}\mathcal{C}\left(m,n,\mathcal{I},\mathcal{S}_\pi,\eta,g\right)\geq\nonumber\\&\sup_{p}\sup_{\gamma}
\Big\{-\frac{1}{2\beta}(\gamma\log{e}-\log{\left(1+\gamma\right)})\nonumber\\&
-\lim_{n\rightarrow\infty}\frac{1}{n}\log{\mathbb E_{\tilde{X},\mathbf{C ,G}}\left(\left(\mathbb E_{b}\left(e^{\frac{-\gamma r^2}{2\left(1+\gamma\right)m}|b^T\mathbf {CG}\tilde X|^2}\right)\right)^m\right)}\Big\},
\end{align}
Now using Varadahn's lemma \cite{Varadhan1}, we compute 
\begin{align}
\log{e}\times\lim_{n\rightarrow\infty}\frac{1}{n}\ln{\mathbb E_{\tilde{X},\mathbf {C,G}}\left(\left(\mathbb E_{b}\left(e^{\frac{-\gamma r^2}{2\left(1+\gamma\right)m}|b^T\mathbf {CG}\tilde X|^2}\right)\right)^m\right)}.
\end{align}
By substituting $r$ by $ \sqrt{\frac{2\eta}{{\sigma_p^2}{\left(\sigma_\pi^2+\mu_\pi^2\right)}{\left(\sigma_c^2+\mu_c^2\right)}}}$ and letting $n\rightarrow\infty$, we obtain

\begin{align}\label{Eq:global4}
&\mathbb E_{b}\left(e^{\frac{-\gamma r^2}{2\left(1+\gamma\right)m}|b^T\mathbf {CG}\tilde X|^2}\right)\approx\nonumber\\& \bigg({1+\frac{2\eta\gamma\beta}{\sigma_p^2(1+\gamma)(\sigma_c^2+\mu_c^2)}}\Big(\frac{1}{n}\sum_{i=1}^{n}(C_iG_i\tilde X_{i})^2\Big)\bigg)^{-\frac{m}{2}}.
\end{align}
Letting 
$f(\theta)=-\frac{1}{2\beta}\ln(1+\frac{2\eta\gamma\beta\theta}{(1+\gamma)(\sigma_c^2+\mu_c^2)(\sigma_\pi^2+\mu_\pi^2)})$ and using Varadahn's lemma, we get
\begin{align}\label{Eq:global5}
&\log{e}\times\lim_{n\rightarrow\infty}\frac{1}{n}\ln{\mathbb E_{\tilde{X},\mathbf{ C,G}}\left(\left(\mathbb E_{b}\left(e^{\frac{-\gamma r^2}{2\left(1+\gamma\right)m}|b^T\mathbf {CG}\tilde X|^2}\right)\right)^m\right)}\nonumber\\&=\log{e}\times\lim_{n\rightarrow\infty}\frac{1}{n}\ln{\mathbb E_{\tilde{X},\mathbf {C,G}}}(e^{n f(\frac{1}{n}\sum_{i=1}^{n}(C_iG_i\tilde X_{i})^2)})\nonumber\\ &\log{e} \times\sup_{\theta\in \mathbb R}\{f(\theta)-I(\theta)\}.
\end{align}
Thus, the desired result follows. Also for a complex channel, the following lower bound holds true
\begin{align}
&\zeta(\beta,\mathcal I,\mathcal S_\pi,\eta ,g,c,\rho=0)\geq \sup_{p}\sup_{\gamma}
\Big\{-\frac{1}{\beta}(\gamma\log{e}-\log{\left(1+\gamma\right)})\nonumber\\&-\log{e}\times\nonumber\\&\sup_{\theta\in \mathbb R^3}\{-\frac{1}{2\beta}\ln\Big(1+\frac{2\eta\gamma\beta}{\sigma_p^2(1+\gamma)(\sigma_c^2+\mu_c^2)(\sigma_\pi^2+\mu_\pi^2)}(\theta_1+\theta_2)\nonumber\\
&+\big(\frac{2\eta\gamma\beta}{\sigma_p^2(1+\gamma)(\sigma_c^2+\mu_c^2)(\sigma_\pi^2+\mu_\pi^2)}\big)^2(\theta_1\theta_2-\theta_3^2)\Big)-I(\theta)\}\Big\},
\end{align}
where $I(\theta)$ is the Legendre transform of $\Big(\text{Re}(b_1C_1G_1\tilde X_1)^2,\text{Im}(b_1C_1G_1\tilde X_1)^2,\text{Re}(b_1C_1G_1\tilde X_1)\text{Im}(b_1C_1G_1\tilde X_1)\Big)$, where $b_1$ is the first entry of $b$.\\\hfill$\Box$

\subsection{Upper Bound for the Sum Capacity of CDMA Systems with Perfect Power Estimation}\label{ProofUpperPerfComp}

If the input alphabets come from a finite set, we have the following upper bound for the average sum capacity 
\begin{align}
&\zeta(\beta,\mathcal  I ,\mathcal S_\pi,\eta , g,c,\rho=0)\nonumber\\&\leq  \min\Big(\log{|\mathcal I|},\frac{1}{2\beta}\max_{p(\cdot)} \log (1+\beta\frac{\text{Var}[ A_1C_1G_1X_1]}{\sigma^2})\Big),
\end{align}
where $A_1$, $C_1$, $G_1$, $X_1$ and $N_1$ are independent random variables with distributions $\pi(\cdot)$, $c(\cdot)$, $g(\cdot)$, $p(\cdot)$ and $\mathcal N\left(0,\sigma^{2}\right)$, respectively.

From Theorem \ref{Th:FiniteUpper} and the central limit theorem, when $n,m \rightarrow \infty$ and $\frac{n}{m}\rightarrow\beta$, $Y_1$ is a complex Gaussian random variable with a covariance matrix of $\Sigma$. Hence, for a fixed distribution $p\left(\cdot\right)$, 
\begin{eqnarray}
\zeta(\beta,\mathcal I,\mathcal S_\pi,\eta,g,c,\rho=0)\leq \frac{1}{\beta}\log\sqrt{\frac{\det~\Sigma}{\sigma^4}}
=\frac{1}{2\beta}\log\frac{\det~\Sigma}{\sigma^4}.
\end{eqnarray}
Maximizing over all distributions $p(\cdot)$, one can get the second term.

In such a system, the sum capacity is upper bounded by
\begin{align}\label{Eq:kasraupper}
&\zeta(\beta,\mathcal  I ,\mathcal S_\pi,\eta,g,c,\rho=0)\leq \nonumber \\  &\min\Bigg(\log{|\mathcal I|},\frac{1}{2\beta}\max_{p\left(\cdot\right)} \log {\left(\frac{\det~\Sigma}{\sigma^4}\right)}\Bigg),
\end{align}
where $\Sigma$ is the covariance matrix of real and imaginary parts of $\sqrt{\beta}a_1X_1+N_1$, in which $a_1$ and $X_1$ are two independent random variables with the corresponding distributions $\pi(\cdot)$ and $p\left(\cdot\right)$; $N_1$ is a complex Gaussian random variable with independent real and imaginary with variance of $\sigma^2$.
\hfill$\Box$

\section*{Acknowledgment}
We are sincerely grateful to Dr. K. Alishahi and Mr. M. Mansouri for their helpful comments.

\bibliographystyle{IEEEtran}
\bibliography{Journal}
\end{document}